\documentstyle[preprint,aps]{revtex}
\input epsf.tex
\def\DESepsf(#1 width #2){\epsfxsize=#2 \epsfbox{#1}}
\begin{document}

\preprint{\vbox{\hbox{OITS-600}\hbox{UCRHEP-T162}\hbox{}}}
\draft
\title {Towards a viable grand unified model with $M_G\sim M_{\rm string}$ and
$M_I\sim 10^{12}$ GeV}
\author{\bf N. G. Deshpande $^{\dagger} $, B. Dutta$^{\dagger} $, and E.  Keith
$^{\ast} $ }\address{$^{\dagger} $ Institute of Theoretical Science, University
of Oregon, Eugene, OR 97403\\$^{\ast} $ Department of Physics, University of
California, Riverside, CA 92521}
\date{April, 1996}
\maketitle
\begin{abstract} We present a model based on the gauge group
SU(2)$_L\times$SU(2)$_R\times$SU(4)$_C$ with  gauge couplings that are found to
be   unified at a scale $M_G$  near the string unification scale.
   The model breaks to the minimal  supersymmetric standard model at a scale
$M_I\sim 10^{12}$ GeV, which is instrumental in producing a neutrino in a mass
range that can serve as hot dark matter and this scale can also solve the strong
CP problem via the Peccei-Quinn (PQ) mechanism with an invisible harmless axion.
 We show how this model can accommodate
 low and high values of $\tan\beta$ and ``exotic" representations that  often
occur in string derived models. We show that this model has lepton flavor
violation which can lead to processes which are one or two  orders of magnitude
below the current experimental limits.
\end{abstract}

\newpage  The conventional scale of  supersymmetric grand unification is taken
to be
$M_G\sim 2\cdot10^{16}$ GeV, because this is where the MSSM gauge couplings are
found to converge if one assumes a ``dessert" between  about 1 TeV and   and
that scale. However, in superstring theory the unification point  is not  a free
parameter but is predicted to be a function of the gauge coupling at  that scale
in the $\overline{\rm MS}$ scheme as follows \cite{[kap]}:
\begin{eqnarray} M_{\rm string}\approx 7 g_{\rm string}\cdot 10^{17}\, {\rm
GeV}\, ,
\end{eqnarray} which predicts $M_{\rm string}$ to be approximately 25 times
greater  than  the conventional value of $M_G$.  Stringy threshold effects have
not yet proven to be at all useful in closing the gap between
$M_G$ and
$M_{\rm string}$ in any realistic string model \cite{[die]}, and neither have
weak to  TeV scale MSSM thresholds.   At present, it is not clear if string
grand unified theories (GUTs) or models that have  non-standard
 Kac-moody levels   and hence non-conventional hypercharge normalizations may
one day be able to rectify the situation \cite{[die]}.  However, it has been
shown that extra non-MSSM matter that appears in some realistic  string models
can lead to a successful raising of the
 unification scale \cite{[die]}. One obvious and attractive approach to adding
extra matter, ``populating the dessert,"  would be to add an intermediate gauge
symmetry. Such realistic string models  have  been built for cases where the
intermediate symmetry is the flipped SU(5)$\times$ U(1) \cite{[SU5]} or
SO(4)$\times$ SO(6)$\sim$ SU(2)$_L\times$SU(2)$_R\times$SU(4)$_C$
\cite{[AL],[ALR],[AM],[L]},  and sometimes the field content of these models
have been found to alleviate the discrepancy between the string and gauge
unification scales \cite{[AM],[L],[SF]}. In this letter, we shall investigate
what field content and additional
 constraints may be required to have a model with gauge unification at the
string scale, that has SU(2)$_L\times$SU(2)$_R\times$SU(4)$_C$ breaking to the
MSSM at an intermediate scale of $\sim 10^{12}$ GeV,  that allows a PQ symmetry
also  broken $\sim 10^{12}$ GeV with a weakly coupled axion, and where the scale
of gauge symmetry breaking and the  PQ symmetry breaking are determined by a
single parameter in the superpotential. We will also discuss the  constraint on
the  field content necessary to obtain, if one desires, a low value for
$\tan\beta$.

An intermediate SU(2)$_R\times$SU(4)$_C$ breaking scale of order $10^{12}$ GeV
is very attractive for  two reasons: (1) if the B-L gauge symmetry is broken at
around $10^{11}$-$10^{12}$ GeV, one can easily  get a neutrino mass in the
interesting range of of about 3-10 eV, making it a candidate for the hot dark
matter
\cite{[SS]}, and (2) if the strong CP problem is solved via the Peccei-Quinn
(PQ) mechanism, this PQ symmetry is required to be broken approximately within
the above window so that  the axion has properties which are consistent with the
lack of observation up to now and the cosmological constraints\cite{[axion]}. As
for the scale  at which the hypothetical PQ-symmetry is broken, perhaps the most
elegant possibility is if it is tied in with the breaking of an intermediate
gauge symmetry, so that there is only one scale between the weak and string
scale to be explained. To obtain the
$\tau$-neutrino mass in the interesting eV range without an intermediate gauge 
symmetry breaking scale
 one has to use a method that either involves a carefully chosen Yukawa coupling
to an
 SU(2)$_R$ triplet, which only arises for particular non-standard Kac-Moody
levels, or non-renormalizable operators with SU(2)$_R$ doublets
\cite{[ML10]}.
 Unlike in the case of an intermediate scale, both these methods require
abandoning the attractice
$b-\tau$ unification hypothesis except in the case of the SU(2)$_R$ doublets
and  high
$\tan\beta \sim m_t/m_b$ which requires greater tuning of the Higgs potential
parameters and may also require $M_G$ scale D-terms.     

 How the  one loop MSSM  beta functions have to be changed at an intermediate
scale to increase the scale of  gauge unification has been studied in a recent
paper\cite{[MR]}.  It is found that there exist only a few acceptable solutions
with a single  cleanly defined intermediate scale far below the unification
scale. In fact if  we demand an intermediate scale as mentioned in the above
paragraph, the  necessary relative  changes in the beta functions of the MSSM
are given as follows: 
\begin{eqnarray}
\Delta b_2-\Delta b_1=2\, ,\, \Delta b_3-\Delta b_2=1\, ,
\end{eqnarray} where the hypercharge has been normalized in the standard GUT
manner and  $b_i=-2 \pi \partial\alpha_i^{-1}/\partial\ln\mu
$.

In this paper the additional field content we choose at the scale $M_I$ is as
follows: the additional vector representation fields necessary to complete the
SU(2)$_L\times$SU(2)$_R\times$SU(4)$_C$ symmetry, 2 copies of the chiral fields
$H=(1,2,4)\equiv ({\bar u}_H^c,{\bar d}_H^c,{\bar E}_H^c,{\bar  N}_H^c)$ and
${\bar H}=(1,2,{\bar 4})\equiv ({ u}_H^c,{ d}_H^c,{ E}_H^c,{  N}_H^c)$, and
chiral singlets $S=(1,1,1)$ which are necessary for the right-handed neutrinos
$N^c_i$ to acquire large Majorana masses. We also add a chiral field
$D=(1,1,6)$ to make all the non-MSSM Higgs modes massive along with two copies 
of chiral fields $\Phi = (2,2,1)$, which contain the MSSM Higgs. There are  of
course the usual 3 MSSM matter generations that include right-handed neutrinos 
$F=(2,1,4)\equiv (u,d,\nu ,e)$ and ${\bar F}=(1,2,{\bar 4})\equiv
(u^c,d^c,N^c,e^c)$. The SU(2)$_R\times$SU(4)$_C$ gauge symmetry is broken to the
U(1)$_Y\times$SU(3)$_c$ by
$\left< H\right> =\left< {\bar N}_H^c\right> \, ,\, \left< {\bar H}\right>
=\left< N_H^c\right> \sim M_I$. This causes 9 of the 21 gauge fields to  become
massive.
 The fields in H and ${\bar H}$ which combine with the corresponding components 
of the gauge fields are $u_H^c$, ${\bar u}_H^c$, $E_H^c$,
${\bar E}_H^c$, and a linear combination of $N_H^c$ and ${\bar N}_H^c$  
orthogonal to that of hypercharge to make super  gauge-Higgs multiplets with a
common mass of the order of $M_I$. In such a case, it is easy to see that  any
number of copies of $H\, ,\, {\bar H}$
 that might exist and get VEVs of order $M_I$ would not leave any massless modes
except for the ones  corresponding to the hypercharge generator since all of the
Higgsinos corresponding to these modes acquire mass through
gaugino-Higgsino-$<$Higgs$>$ terms. The linear combination of 
$N_H^c$ and ${\bar N}_H^c$ corresponding to the hypercharge generator 
 gets mass of the order of $M_I$ through the terms 
$\lambda_{H{\bar H}S} H{\bar H}S$. Note that we do not add terms like 
$M_{H{\bar H}}H{\bar H}$ in the superpotential which cause the breaking of SUSY  
 via F-terms. The presence of R symmetry\cite{[WB]} and the non-existence of a
VEV for S could,  for example, forbid these terms.  The chiral fields $d_H$ and
${\bar d}_H^c$ in these  representations become massive with the  help of the
field $D=(1,1,6)\equiv (d_D^c,{\bar d}_D^c)$. This causes
$d_H$,
${\bar d}_H^c$, $d_D^c$, and ${\bar d}_D^c$ to all get mass of order $M_I$
through the superpotential terms $\lambda_{HHD}HHD$ and
$\lambda_{{\bar H}{\bar H}D}{\bar H}{\bar H}D$ when $H$ and ${\bar H}$ gets
VEVS. Note that to avoid rapid proton decay, we also need to impose a symmetry
on the superpotential (for example PQ symmetry as discussed later)
  that forbids terms of the type
$FFD$ and ${\bar F}{\bar F}D$ unless the couplings are extremely small.

 The existence of the field D and S are crucial to make all the Higgs modes
massive. As a matter of fact, in a previous paper Ref. \cite{[AM]} the field 
content without D and one of the bidoublets have been used to raise the
unification  scale. This field content also satisfies Eq.(2). (Note that in
SO(10): 
${10}\rightarrow (1,1,6)+(2,2,1)$.)  However the choice of this minimal field
content is problematic in  practice as we have already pointed out that all the
non-MSSM Higgs modes do not become massive at the breaking scale
$M_I$, and hence the  gauge coupling renormalization group equations (RGEs) are
modified beneath the scale $M_I$. This model also suggests complete third
generation Yukawa coupling unification at the intermediate scale, and hence 
requires  large
$\tan\beta$, due to the existance of only one bidoublet Higgs.
 Instead of the second bidoublet,
 we could have 2 copies of fields transforming as
$(2,1,1)+(1,2,1)$. As a matter of fact, in string derivations of
SU(2)$_L\times$SU(2)$_R\times$SU(4)$_C$ models the exotic representations
$(2,1,1)$,
$(1,2,1)$, and $(1,1,4)+(1,1,{\bar{4}})$ tend to occur. Since the field content
without D and S satisfy Eq. (2), the constraints on  the additional field
content is given by :
\begin{eqnarray} n_D\geq 1\, ,\, n_D+n_4=(n_\Phi - 1) + {1\over 2}n_2\, ,
\end{eqnarray}    where $n_D$ is the number of copies of fields transforming as
$(1,1,6)$, $n_\Phi$ is the number of fields transforming as $(2,2,1)$, $n_4$ is
the number of copies of $(1,1,4)+(1,1,{\bar 4})$, and $n_2$ is the number of
copies of $(2,1,1)+(1,2,1)$, which are all to be given mass of order $M_I$. Of
course, Eq. (2) gives no constraint on $n_{S}$, the number of copies of singlet
$(1,1,1)$ fields. It may be of interest to note that, for example, the first  
string derived version of the model found in Ref. \cite{[AL]} has 
$n_D=4$, 
$n_4=1$, 
$n_\Phi=4$, 
$n_2=10$, along with the necessary  2 copies of $H+{\bar H}$ ($N_H=2$)  at the
string scale and 3 generations of  $F+{\bar F}$ $(N_F =3)$, as well as several
SU(2)$_L\times$SU(2)$_R\times$SU(4)$_C$ singlets. Some of these
SU(2)$_L\times$SU(2)$_R\times$SU(4)$_C$ may acquire VEVs near the string scale
which can break additional U(1) symmetries and may make some fields super heavy. 

We now discuss how a low $\tan\beta$ scenario may be implimented in the model.
To allow for the possibility of low
$\tan\beta$, one needs the MSSM Higgs doublets $\phi_u$ and $\phi_d$ to not come
primarily from the same bidoublet $\Phi_i$. If we assume two bidoublets 
$\Phi_1$ and
$\Phi_2$, no difficulty would exist in the breaking of the
SU(2)$_L\times$SU(2)$_R\times$SU(4)$_C$ model to the MSSM if  one bidoublet was
to remain light at
$M_I$ and the other was to be given mass of order $M_I$. However here we want
one linear combination, from the two bidoublets, of down (or up) type Higgs
superfield SU(2)$_L$ doublets to remain massless and the other combination to
have mass of order
$M_I$. (Otherwise, we would spoil the gauge coupling RGE analysis, which is the
primary motivation for this model.) This can be accomplished  through a
modification of a method that has been used in conventional SO(10)  GUTs.
\cite{[BaMoha]}.
 Consider adding to the model a pair of fields $H_L$ and ${\bar H}_L$
transforming as
$(2,1,4)$ and $(2,1,{\bar 4})$, respectively, and  also increasing the number of
$H,{\bar H}$ pairs to be $N_H=3$ so that  Eqn.(2) is still satisfied. For
simplicity, we discuss the general case and  will not refer to specific choices
of any PQ charges for the additional field content of this paragraph.  If 
explicit mass terms for these fields are originally forbidden, they can be
generated at the intermediate scale to be of order
$M_I$ through terms such as $\lambda_{H_L} H_L{\bar H}_LS_i$ where $S_i$ gets a
VEV at the scale
$M_I$ in the manner as previously discussed. In the limit of neglecting weak
scale masses, the existence of the  superpotential terms
$W_D=\lambda_1 {\bar H}H_L\phi_1 + \lambda_2 {H}{\bar H}_L\phi_2$ and a symmetry
(for example, the superpotential can be assumed to be invariant under the
transformation: $ H_L=-H_L$, $\phi_1=-\phi_1$ and $S_i=-S_i$)  forbidding the 
terms $\lambda_1 {\bar H}_LH\phi_1 + \lambda_2{H}_L{\bar H}\phi_2$ 
 would lead to the following form for the mass matrix for the SU(2)$_L$ doublets:
\begin{eqnarray} M_D &=& \left(\matrix{\lambda_{H_L} v_{S_i} & \lambda_1 v_H & 0
\cr
\lambda_2 v_{\bar H} & 0 & 0 \cr 0 & 0 & 0}\right) \, ,
\end{eqnarray}  written in a basis where the rows stand for $({\bar
H}_{L_{\phi_u}},\phi_{1u} ,\phi_{2u} )$ and the columns for
$(H_{L_{\phi_d}} ,\phi_{1d} ,\phi_{2d} )$ in obvious notation and all VEVs are
of order $M_I$. This matrix naturally will have two large eigenvalues of order
$M_I$, and has one massless eigenvalue which is composed of
$\phi_{1d}=\phi_d$ and
$\phi_{2u}=\phi_u$ and serves as the MSSM Higgs. Note that had D-parity not been
broken in the model, we would not have had to have added any additional field
content.

The gauge couplings in this model have the following one-loop beta functions:
\begin{eqnarray} b^{224}_{i}=\left( \matrix{ -6\cr -6\cr -12\cr} \right) +n_F
\left( \matrix{  2\cr 2\cr 2\cr} \right)  +n_{\Phi} \left( \matrix{  1\cr 1\cr
1\cr} \right) +n_{D} \left( \matrix{  1\cr 1\cr 1\cr} \right) +N_{H_L} \left(
\matrix{  0\cr 4\cr 2\cr} \right) +N_{H} \left( \matrix{  4\cr 0\cr 2\cr}
\right) \ \ \, ,
\end{eqnarray} and the following two-loop beta functions:
\begin{eqnarray} b^{224}_{ij}&=&\left( \matrix{ -24&0&0\cr 0&-24&0\cr
0&0&-96\cr} \right)   + N_F \left( \matrix{ 14&0&15\cr  0&{14}&15\cr
3&3&31\cr}\right) + n_{\Phi} \left( \matrix{  7&3&{0}\cr 3&{7}&{0}\cr
{0}&{0}&{0}\cr} \right)+ n_{D} \left( \matrix{  0&0&{0}\cr 0&{0}&{0}\cr
{0}&{0}&{18}\cr} \right)\\\nonumber &&+N_{H_L} \left( \matrix{  0&0&{0}\cr
0&{28}&{30}\cr {0}&{6}&{31}\cr} \right) + N_{H} \left( \matrix{  28&0&{30}\cr
0&{0}&{0}\cr {6}&{0}&{31}\cr} \right) \, , 
\end{eqnarray} where $i=SU(2)_R,SU(2)_L,SU(4)_C$  respectively in the matrices,
$N_{H_L}$ is the number of 
 field pairs(to be used for the low $\tan\beta$ scenario) transforming as
$\left(2,1,4\right)+\left(2,1,{\bar 4}\right)$, $N_F=3$ always,  and we have
left out the contributions from exotic representations since they can be easily
calculated and we will not use them in our examples here. In Table 1, we show
some sample gauge coupling unification results for the model in the case of
$n_\Phi =2$,
$n_D=1$ and 
$N_{H}=2$ which requires complete third generation Yukawa coupling unification 
at the scale $M_I$ and which
 we refer to as the high $\tan\beta$ scenario, for different values of
$\alpha_s (M_Z)$ and the effective SUSY scale $M_S$ with $
\sin^2{\theta_W} =0.2321$,
$\alpha (M_Z)=1/127.9$, $M_I=10^{12}$ GeV, and the top quark mass 
$m_t\approx 180$ GeV. For the case that does not require $\lambda_t=\lambda_b$
at $M_I$, which we refer to as the low 
$\tan\beta$ scenerio, we use $n_\Phi =2$,
$n_D=1$, $N_H=3$ and $N_{H_L}=1$. We display
$M_G$, which we compare with the string scale prediction from Eq. (1). We
 do not include string threshold effects as we do not know the entire field
content near the string scale, however we note from Ref.
\cite{[die]} that the effect of these thresholds in this model should in general
tend towards having the bennificial effect of further reducing the small
discrepancy between $M_G$ and $M_{\rm string}$.

We note that the appearance of the intermediate breaking scale can occur through
a single parameter in the singlet sector of the model. For example, consider an
R symmetry  invariant superpotential
\begin{eqnarray}  W&=&\left(\sum_{i,j=1}^{2} \lambda_{ij} H_i{\bar H}_j -r\right)
S_0 +...,
\end{eqnarray} where $r$ is of order $M_I$ and $S_0$ has no VEV. It would then
be the most natural case for the VEVs acquired by all four fields to be of
similar order. This mechanism can easily be extended to link the breaking  of a
PQ-symmetry  with the breaking of the intermediate gauge symmetry. For example,
suppose a PQ-symmetry exists with $F,{\bar F}$ having a PQ charge of 1,
bidoublet(s)
$\Phi_i$ which contain the MSSM Higgs doublets and has (have) PQ charge -2, that
the fields
$H,{\bar H}$ have no or opposite PQ charges,  singlets
$S_0,S_{2},S_{-2}$ exist with the subscript denoting the PQ charge,  and that
R-symmetry prevents
$M_{2,-2}S_2S_{-2}$ mass term from existing in the superpotential, then the
following superpotential is possible:
\begin{eqnarray} W={\lambda_{\Phi_{ij}}\over M_{\rm Pl}}\Phi_i\Phi_j
S_2S_2+\left(\lambda_S S_{-2}S_2 +
\sum_{i,j=1}^{2} \lambda_{ij} H_i{\bar H}_j -r\right) S_0+...
\end{eqnarray} where we have allowed a non-renormalizable term
\cite{[chun]} so as to not need to  fine-tune the $\mu$ parameter to a very
large value or
$\lambda_{\Phi_{ij}}$ to a tiny value. We observe that we can choose the  PQ
charge of $D$ to be 0 for example so as to forbid terms like $FFD$ and 
${\bar F}{\bar F}D$ that would cause the rapid proton decay.

We note that in the model we are discussing there are no SU(2)$_R$ triplet
fields, therefore one must rely on an extended version of the seesaw mechanism 
 \cite{[LeeMoha]}. This mechanism in this  scenario is described by the following
$3\times 3$ mass matrix for $\nu_a$, $N_a^c$, and singlets $S_a$ with $a=1,2,3$:
\begin{eqnarray} M_{\nu}&=&\left(\matrix{
               0                &\lambda_{(u)}v_u      &0     \cr
               \lambda^T_{(u)}v_u       &0               &f v_I \cr
               0                &f^T v_I        &M   }\right)
\end{eqnarray}  where $\lambda_{(u)}$, $f$, and $M$ are $3\times 3$ matrices,
$v_I$ is a VEV of order
$M_I$,  and $v_u$ is the electroweak breaking VEV of the MSSM Higgs doublet
$H_u$. Ignoring intergenerational mixing by pretending $\lambda_{(u)}$, $f$, and
$M$ are diagonal, then the mass of the $a$th light Majorana neutrino is given by
$m_{\nu_a}\sim \left(\lambda_{(u)} v_u\right)^2 M_a/f_a^2v_I^2$. To get back the
usual seesaw relation and have
$m_{\nu_\tau}$  be in the eV range, we need $M_a\sim M_I$. This can be done  by
adding to the field content of the previous paragraph three generations of
singlets transforming as $S_{-1}$ and at least one
$S_1$,  assuming once again from R-symmetry that the explicit mass terms
$M_{S_1,S_{-1}}S_1S_{-1}$ are forbidden in the
SU(2)$_L\times$SU(2)$_R\times$SU(4)$_C$ superpotential, and giving
$H_i,{\bar H}_j$ PQ charges of 0. All this allows the terms
$\lambda_{ij} {\bar F}_i H_j S_{-1}$,
$\lambda_{-1,-1,2} S_{-1}S_{-1}S_2$, and $\lambda_{-1,1,0}S_{-1}S_1S_0$ to be in
the superpotential, which at the intermediate scale  would give
$S_{-1}$ both a mass and a VEV of order $M_I$ and  give the desired size to the
entrees in the seesaw mixing matrix.

Lastly, we consider a signal of the model. Recently, GUT scale physics
 have been shown to be significant sources of lepton flavor violation in SUSY
grand unification with a universal SUSY soft breaking boundary condition
appearing near the reduced Planck scale \cite{[LJ],[AS],[cia]}. However the
parameter  space where these signals could be observed is somewhat constrained
by the  experimental constarints   of $b\rightarrow s\gamma$ \cite{[us]}. More
recently, it has also been shown that even if the universal boundary condition
is taken at the GUT scale, that an intermediate gauge symmetry breaking can also
be a significant source of lepton violation \cite{[dku]}. We now show that the
rate of
$\mu\rightarrow e\gamma$ can be within two orders of magnitude of experiment in
the model discussed in this letter. For some parameter space, it can be above
the experimental limit. In SU(2)$_L\times$SU(2)$_R\times$SU(4)$_C$ gauge
symmetry, the quarks and leptons are unified. Hence, the $\tau$-neutrino Yukawa
coupling is the same as the top Yukawa coupling. Through the RGE's, the effect
of the large $\tau$-neutrino Yukawa coupling is to make the third generation
sleptons lighter than the first two generations, thus mitigating the GIM
cancellation in one-loop leptonic flavor changing processes which involve
virtual sleptons. 

The superpotential terms which will be responsible for giving the SM fermion
masses have the following form :
\begin{eqnarray} W_Y&=&{\bf \lambda_{F_u}}{\bf F}{\bf \Phi_2}{\bf {\bar F}} +
 {\bf\lambda_{F_d}}{\bf F}{\bf \Phi_1}{\bf{\bar F}} , ,
\end{eqnarray} where all group and generation indices have been suppressed.
 $\Phi_1$ and $\Phi_2$ are the two bidoublets. We have assumed that $\Phi_2$
contains the MSSM Higgs doublet which gives masses to the up quarks and Dirac
masses for the neutrinos and
$\Phi_1$ contains the doublet which gives masses to the down quarks and the
charged leptons.   Now we give the RGEs for the soft SUSY breaking parameters 
which we need for the intermediate gauge symmetry . First of all, there are
gaugino masses $M_i$ corresponding to each
$g_i$. Secondly, corresponding to each tri-linear superpotential coupling
${\bf \lambda_i}$ there is a tri-linear scalar term with the coupling ${\bf A_i}
{\bf \lambda_i}$ at $M_G$. Finally there are soft scalar mass terms $m_i^2$
 for each of the the fields
$F_{L,R}$, and $\Phi_{1,2}$.
\begin{eqnarray} {\cal D}\lambda^2_{F_{ug}} &=& -\sum_i{c^{\left(\lambda_F
\right)}_ig^2_i}+
\left(4+4\delta_{g3}\right)\lambda_{F_t}^2,\\ {\cal D}\lambda^2_{F_{dg}} &=&
-\sum_i{c^{\left(\lambda_F \right)}_ig^2_i}+4\delta_{g3}\lambda_{F_t}^2,\\ {\cal
D}M_i&=&b_ig_i^2M_i,\\ {\cal D}A_{F_{ug}} &=&\sum_i{c^{\left(\lambda_F
\right)}_ig^2_iM_i}+
\left(4+4\delta_{g3} \right)\lambda_{F_t}^2A_{F_t},\\ {\cal D}A_{F_{dg}}
&=&\sum_i{c^{\left(\lambda_F \right)}_ig^2_iM_i}+
4\delta_{g3}\lambda_{F_t}^2A_{F_t},\\ {\cal D}m^2_{F,{\bar
F}}&=&-\sum_i{c^{\left(F,{\bar F}\right)}_ig^2_iM_i^2} +
2\lambda_{F_t}^2X\delta_{g3},\\ {\cal
D}m^2_{\Phi_1}&=&-\sum_i{c^{\left(\Phi\right)}_ig^2_iM_i^2},\\ {\cal
D}m^2_{\Phi_2}&=&-\sum_i{c^{\left(\Phi\right)}_ig^2_iM_i^2} + 4\lambda_{F_t}^2X 
,
\end{eqnarray}  where $g$ refers to generation and $i$ refers to the gauge,
\begin{eqnarray}\nonumber c^{(\lambda_F )}=\left( 3,3,{15\over 2}\right)\, ,\,
c^{(F)}&=&\left( 3,0,{15\over 2}\right) , c^{({\bar F})}=\left( 0,3,{15\over
2}\right), c^{(\Phi )}=\left( 3,3,0\right),\\\nonumber  X&\equiv&
m^2_{F}+m^2_{{\bar F}}+M^2_{\Phi_2}+A^2_{F_t}\nonumber,
\end{eqnarray} and we have used
\begin{eqnarray} {\cal D}\equiv {16\pi^2\over 2}{d\over dt}\nonumber,
\end{eqnarray} where $t=\ln{(\mu /{\rm GeV})}$ with $\mu$ being the scale. At
the scale $M_G$, we assume a universal form to the soft SUSY breaking parameters
i.e. all gaugino masses $M_i(M_G) =m_{1\over 2}$, all tri-linear scalar
couplings $A_i(M_G)= A_0$, and all soft scalar masses $m^2_i (M_G)=m_0^2$.  At
the scale $M_I$, we match the intermediate gauge symmetry breaking effective
theory parameters with the MSSM parameters in the usual fashion. We run all the
RGEs according to the MSSM \cite{[SUSYYUK]} down to the top scale. Details of
$\mu\rightarrow e\gamma$ with an intermediate symmetry  are presented in Ref.
\cite{[dku]}.  As an example, we show the specific case of the universal gaugino
masses
$m_{1/2}=145$ GeV and universal tri-linear soft breaking parameter $A_0=0$ at the
unification scale, $\alpha_s (M_Z)=0.119$, $M_G=10^{17.87}$ GeV,
$\alpha_G=1/12.4$,
 $m_t=176$ GeV and
$m_b=4.35$ GeV, and with minimal field content for low $\tan\beta =2$ in Fig. 1.
We have plotted the function \begin{eqnarray} l_r\equiv {\rm Log}_{10}{\left(
{B\over B_{\rm exp}}\right)}\, , \end{eqnarray}  where $B$ is the predicted
$\mu\rightarrow e\gamma$ branching ratio and
$B_{\rm exp} =4.9\cdot 10^{-11}$ being the experimental 90 $\%$ confidence limit
upper bound on the branching ratio. With $A_0=0$, $A_i$ is always negative at
the weak scale. Consequently we find that  $\mu<0$ gives  a greater branching
ratio.

In summary, we have discussed a model that allows gauge coupling unification  in
the vicinity if the string scale and can have an intermediate
SU(2)$_L\times$SU(2)$_R\times$SU(4)$_C$ gauge  symmetry breaking scale $M_I$ of
order $10^{12}$ GeV, which is useful for producing a $\tau$-neutrino with a mass
of a few eV and solving the strong CP problem via a PQ symmetry whose breaking
produces a harmless axion. We have shown that a range of field content is
allowed by the model as given by Eq. (3) to satisfy Eqn. (2)  which predicts
$M_G/M_I \approx 10^{6\pm 2}$ GeV, and we have shown that Eqn. (1) which gives
the string prediction between unification scale mass and gauge couplings are
approximately satisfied for some choices of field content. We have also
discussed what SU(2)$_L\times$SU(2)$_R\times$SU(4)$_C$ singlets and global U(1)
charges may be useful to take advantage of $M_I\sim 10^{12}$ GeV and make a
single parameter in the singlet sector responsible for the scale $M_I$. 

Recently, an intermediate scale at $\sim 10^{12}$ GeV has been advocated
\cite{[TK]} to produce monopoles that would explain the high energy cosmic ray 
spectrum. Since our model factors into U(1) group at $M_I$, it would be natural 
for such monopoles to arise in this model. It may be of interest to see if such
monopoles satisfy the requirements of relic abundance required for the suggested 
mechanism.

We are grateful to K. S. Babu and E. Ma for invaluable comments and discussions.
This work was supported by Department of Energy grants DE-FG06-854ER 40224 and
DE-FG02-94ER 40837.

\newpage

\leftline{{\Large\bf Table captions}}
\begin{itemize}

\item[Table 1~:] {The gauge unification scale $M_G$ and string scale $M_{\rm
string}$ (as predicted in Eqn. (1)) are shown for different values of
$\alpha_s(M_z)$
 and effective SUSY scale $M_S$ in both the low and the high $\tan\beta$ models
described in the text. } 
\end{itemize}

\leftline{{\Large\bf Figure captions}}
\begin{itemize}

\item[Fig. 1~:] {$l_r\equiv Log_{10}{\left( B/B_{\rm exp}\right)}$ is  plotted
as a function of of the universal at $M_G$ scale  soft mass $m_0$.\\  The solid
line corresponds to $\mu >0$ , while the dashed line corresponds to $\mu <0$.
\\ $\lambda_{F_{{t}_G}}=1.25 $ and $m_{1/2}=140$ GeV for both lines}. 

\end{itemize}
\newpage
\begin{center}
\begin{tabular}{|c|c|c|c|c|c|}  \hline
$\tan\beta$&$\alpha_s (M_Z)$& $M_s(GeV)$ &1/$\alpha_G$ &  $M_G(GeV)$ &$M_{\rm
string}(GeV)$ \\\hline
 high&0.1258&175&$20.34$&$10^{18.26}$& $10^{17.74}$\\
high&0.1187&$10^3$&$22.10$&$10^{17.92}$& $10^{17.72}$\\\hline
low&0.1192&$175$&$10.65$&$10^{17.83}$& $10^{17.84}$\\
low&0.1144&$10^3$&$12.85$&$10^{18.05}$& $10^{17.88}$\\\hline
\end{tabular}
\end{center}
\newpage

\vfill
\begin{figure}[htb]
\centerline{ \DESepsf(paper4.epsf width 15 cm) } \smallskip \nonumber
\end{figure}

\newpage
\begin{thebibliography}{[001]}
\bibitem{[kap]} V. A. Kaplunovsky, Nucl. Phys \underline {B307} 145, (1988);
ERRATUM-ibid.\underline{B382} 436, (1992).

\bibitem{[die]}   K. R. Dienes and) A. E. Faraggi, Phys. Rev. Lett.
\underline{75} 2646, (1995);
 K. R. Dienes and A. E. Faraggi, Nucl. Phys. \underline{B457} 409, (1995);
 K. R. Dienes, hep-th/9602045; 


\bibitem{[SU5]}  I. Antoniadis, J. Ellis, J.. S. Hagelin and D. V. Nanopoulos,
 Phys. Lett. \underline {B194} 231, (1987); ibid. \underline{B205} 459, (1988),
ibid. \underline {B208} 209, (1988).

\bibitem{[AL]}  I. Antoniadis and G.K. Leontaris, Phys. Lett.\underline {B216}
333, (1989).

\bibitem{[ALR]}  I. Antoniadis , G.K. Leontaris and J. Rizos , Phys. Lett.
\underline {B245} 161, (1990).

\bibitem{[AM]} A. Murayama and A. Toon, Phys. Lett. \underline {B318} 298,
(1993).

\bibitem{[L]}G.K. Leontaris and N. D. Tracas hep-ph/9511280  .

\bibitem{[SF]} S.F. King,  Phys. Lett. \underline{B325} 129, (1994).

\bibitem{[SS]}R. Shafer and Q. Shafi, Nature (London) \underline{359}, 199 
(1992).

\bibitem{[axion]}M. Dine, W. Fischler and  M. Srednicki, Physics lett.
\underline {B104} 199 (1981).

\bibitem{[ML10]}F. Vissani and A. Y. Smirnov, Physics Lett. \underline{B341} 173
(1994); A. Brignole, H. Murayama, and R. Rattazzi,  Physics Lett.
\underline{B335} 345 (1994); D-G. Lee and R. N. Mohapatra, Phys. Rev.
\underline{D52} 4125, (1995).

\bibitem{[MR]} S. P. Martin and P. Ramond, Phys. Rev. \underline{D51} 6515,
(1995).

\bibitem{[WB]}J. Wess and J. Bagger, Supersymmetry and Supergravity, (Princeton
university Press, 1983).

\bibitem{[BaMoha]} K. S. Babu and R. N. Mohapatra, Phys. Rev. Lett.
\underline{74} 2418, (1995).

\bibitem{[chun]} E. J. Chun, Phys. Lett. \underline{B348} 111, (1995).

\bibitem{[LeeMoha]} R. N. Mohapatra,  Phys. Rev. Lett.
\underline{56} 561, (1986), R. N. Mohapatra and J. W. F. Valle, Phys. Rev
\underline{D34} 1642, (1986), D-G Lee, R. N. Mohapatra,  Phys.  Rev.
\underline{D52} 4125, (1995).

\bibitem{[LJ]} R. Barbieri and L. J. Hall, Phys. Lett. \underline{B338} 212,
(1994).

\bibitem{[AS]} R. Barbieri, L. J. Hall, and A. Strumia,  Nucl. Phys
\underline{B445} 219, (1995).

\bibitem{[cia]}P. Ciafaloni, A. Romanino, and A. Strumia, IFUP-TH-42-95.

\bibitem{[us]} T. V. Duong, B. Dutta and E. Keith, OITS-591, UCRHEP-T154, hep-ph
9510441 (to appear in Phys. Lett. \underline{B}).

\bibitem{[SUSYYUK]}For example, see: V. Barger, M. Berger, and P. Ohmann,
Phys.Rev. 
\underline{D47} 1093, (1993); V. Barger, M. Berger, and P. Ohmann, Phys. Rev.
\underline{D49} 4908, (1994).


\bibitem{[dku]}N. G. Deshpande, B. Dutta and E. Keith, hep-ph/9512398.

\bibitem{[TK]}T. W. Kephart and T. J. Weiler, astro-ph/9505134.






\end{thebibliography}
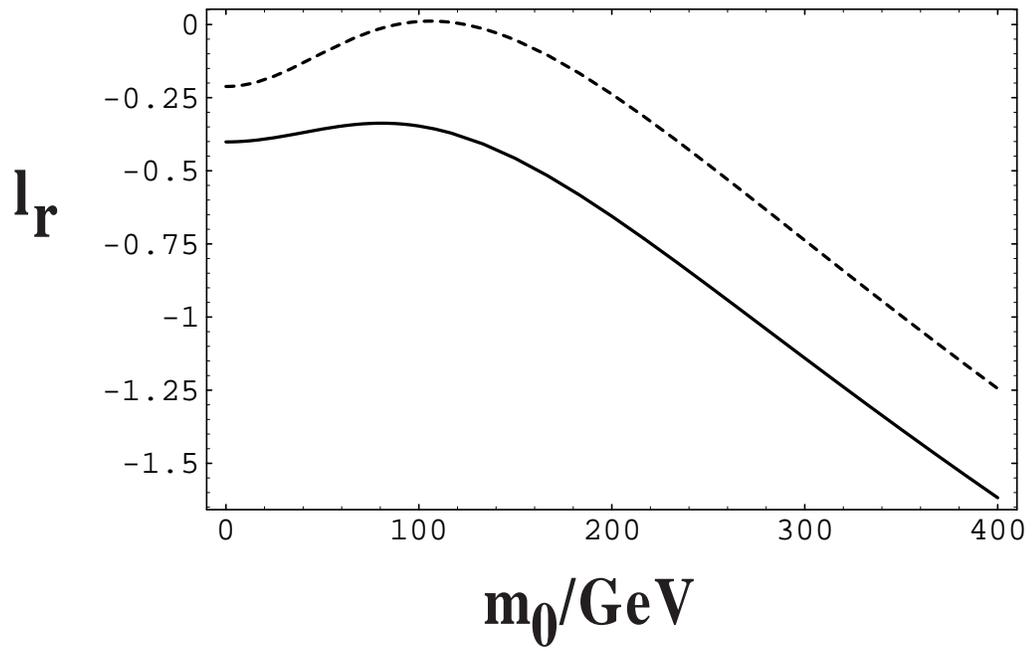
\end{document}